\documentclass[twocolumn,superscriptaddress,floatfix,prb]{revtex4-1}
\usepackage{graphicx,amsfonts,amssymb,amsmath,hyperref,hypcap,enumerate}
\usepackage{color}
\usepackage{cleveref}
\usepackage{resizegather}

\newcommand{\qq}{\boldsymbol{q}}
\newcommand{\QQ}{\boldsymbol{Q}}
\newcommand{\RR}{\boldsymbol{R}}
\newcommand{\rr}{\boldsymbol{r}}

\newcommand{\kk}{\boldsymbol{k}}

\newcommand{\bb}{\boldsymbol{b}}

\begin{document}
\title{Ferromagnetism and superconductivity in twisted double bilayer graphene}

\author{Fengcheng Wu}
%\email{wufcheng@umd.edu}
\affiliation{Condensed Matter Theory Center and Joint Quantum Institute, Department of Physics, University of Maryland, College Park, Maryland 20742, USA}
\author{Sankar Das Sarma}
\affiliation{Condensed Matter Theory Center and Joint Quantum Institute, Department of Physics, University of Maryland, College Park, Maryland 20742, USA}

\date{\today}

\begin{abstract}
	We present a theory of competing ferromagnetic and superconducting orders in twisted double bilayer graphene (TDBG). In our theory, ferromagnetism is induced by Coulomb repulsion, while superconductivity with intervalley equal-spin pairing can be mediated by electron-acoustic phonon interactions. We calculate the transition temperatures for ferromagnetism and superconductivity as a function of moir\'e band filling factor, and  find that superconducting domes can appear on both the electron and hole sides of the ferromagnetic insulator at half filling. We show that the ferromagnetic insulating gap has a dome shape dependence on the layer potential difference, which provides an explanation to the experimental observation that the ferromagnetic insulator only develops over a finite range of external displacement field.  We also verify the stability of the half-filled ferromagnetic insulator against two types of collective excitations, i.e., spin magnons and valley magnons.
\end{abstract}

\maketitle

\section{introduction}
Moir\'e superlattices form in van der Waals bilayers with a small orientation misalignment and/or lattice constant mismatch. Recently moir\'e bilayers have emerged as a platform to study fundamental physics of strongly interacting systems, in view of the discovery of correlated insulating and superconducting states in twisted bilayer graphene\cite{Cao2018Super,Cao2018Magnetic}. Moir\'e superlattices often generate spatial confinement for low-energy electrons, suppress electron kinetic energy, and therefore effectively enhance interaction effects. Evidences of  correlated insulating and superconducting states have so far been reported in three graphene-based moir\'e systems, including twisted bilayer graphene \cite{Cao2018Super,Cao2018Magnetic, Dean2018tuning,kerelsky2018magic,choi2019imaging,sharpe2019emergent,MIT2018_rho,Columbia2018_rho,codecido2019correlated,lu2019superconductors,tomarken2019electronic}, twisted double bilayer graphene\cite{shen2019observation,liu2019spin,cao2019electric,Burg2019,he2020tunable}, and ABC trialyer graphene on hexagonal boron nitride\cite{Chen2019Mott,chen2019signatures,chen2019tunable}
.

Twisted bilayer graphene (TBG) is a subject under intense theoretical study\cite{Balents2018,Senthil2018,Koshino2018, Kang2018,  Dodaro12018, Padhi2018, Guo2018,Fidrysiak2018,  Kennes2018strong,  Liu2018chiral,Isobe2018, You2018, Tang2019, rademaker2018charge,PALee2018, guinea2018electrostatic,Carr2018, Thomson2018, gonzalez2018kohn, Lin2018,Lado2018,Vishwanath2018origin, Ahn2018failure,Bernevig2018Topology,lian2018landau, hejazi2018multiple, liu2018complete, sherkunov2018novel,Venderbos2018,KoziiNematic2019,kang2018strong,xie2018nature,lin2019chiral,bultinck2019anomalous,  Alidoust2019, Angeli2019, Heikkila2018, Lian2018twisted, choi2018electron,Wu2018phonon, wu2019phonon,wu2018topological,wu2019identification}, but the exact nature of the correlated insulating (CI) and superconducting (SC) states in TBG remains unsettled. The  half-filled correlated insulator in TBG crosses over to a metallic state by a strong perpendicular or parallel magnetic field \cite{Cao2018Super,Cao2018Magnetic}, which possibly rules out spin-polarized ferromagnetic states, but leaves a large number of possible non-FM states as candidates, e.g., valley polarized state, and charge/spin/valley density wave states to name a few.

By contrast, there appears to be good experimental evidence that the half-filled CI in twisted double bilayer graphene (TDBG) with  a twist angle $\theta$ around   $1.3^{\circ}$ is ferromagnetic, because the correlation driven insulating gap has been found to be enhanced by an in-plane magnetic field \cite{shen2019observation,liu2019spin,cao2019electric,Burg2019,he2020tunable}.  Possible signature of SC domes in adjacent to the CIs has also been reported in TDBG \cite{shen2019observation,liu2019spin}. These experimental discoveries identify TDBG as another important moir\'e system with strong interaction effects.  Moreover, TDBG represents a simpler as well as a more tunable system compared to TBG, because moir\'e bands of TDBG can be controlled by an out-of-plane electric displacement field and its first moir\'e conduction band can be energetically isolated from neighboring bands, whereas the first moir\'e valence and conduction bands in TBG are typically connected via Dirac points.

In this paper, we theoretically study TDBG FM and SC orders in its first moir\'e conduction band. In our theory, ferromagnetism is driven by Coulomb repulsion as in Stoner model, but superconductivity is mediated by electron-phonon interactions.  FMCI can occur at half filling when spin majority and minority bands are separated in energy by Coulomb exchange interaction. We find that the FM insulating gap is tunable by a layer potential difference $U$ that is generated by an external out-of-plane displacement field. This tunability originates from the strong dependence of the moir\'e bands on $U$,  and agrees with experimental observations \cite{shen2019observation,liu2019spin,cao2019electric,Burg2019,he2020tunable}. We also calculate spin and valley magnon spectrum and verify the stability of the FMCI.

Away from half filling, the state is generally metallic, which can be susceptible to superconducting instability at low temperature. Because electron-acoustic phonon interactions in graphene mediate both spin singlet and spin triplet intervalley Cooper pairing\cite{wu2019phonon},  superconductivity  can take place even in the presence of ferromagnetism, as long as the spinless time-reversal symmetry is preserved. We estimate the transition temperatures of ferromagnetism and superconductivity as a function of filling factor, and find that superconducting domes can appear on both sides of the half-filled ferromagnetic insulator. 

Our paper is organized as follows.  We describe the moir\'e Hamiltonian and band structure of TDBG in Section ~\ref{section:moire}. We find that the moir\'e bands are tunable by $U$, and the van Hove singularity in the non-interacting density of states can be tuned from below to above half filling of the first moir\'e conduction bands. This feature allows the  controlling of  many-body physics using the out-of-plane displacement field.   We present theory for ferromagnetism and superconductivity, respectively, in Sections~\ref{section:FM} and \ref{section:SC}, and make a brief conclusion  in Section ~\ref{section:end}.  Some technical details of the theory are given in Appendices \ref{appA} and \ref{appB}.

\begin{figure}[t!]
	\includegraphics[width=1\columnwidth]{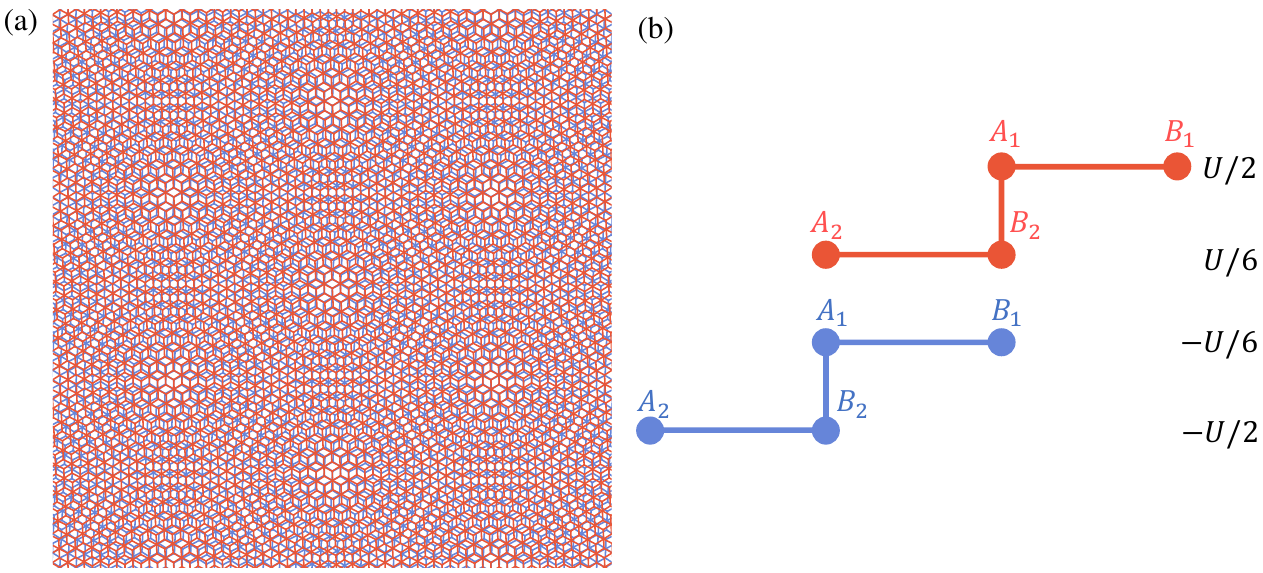}
	\caption{(a) Top and (b) side  view of twisted double bilayer graphene. The top and bottom bilayers are marked by red and blue colors.}
	\label{Fig:moire}
\end{figure}

\section{Moir\'e Bands}
\label{section:moire}

We study TDBG with a small twist angle $\theta$ relative to the AB-AB stacking configuration, and calculate the moir\'e band structure using a continuum   Hamiltonian  generalized from TBG\cite{Bistritzer2011} to TDBG \cite{Zhang2019,chebrolu2019flatbands,Koshino2019,lee2019theory,liu2019quantum, Haddadi2019}.  Within the continuum approximation, $\pm K$ valleys are treated separately. For each AB bilayer graphene, we use the following $k.p$ Hamiltonian in $+K$ valley
\begin{equation}
\mathcal{H}_0(\kk) = \begin{pmatrix}
\Delta_0       & \hbar v_1 k_- & \hbar v_2 k_+ & \gamma_1 \\
\hbar v_1 k_+  &  0            & \hbar v_3 k_- & \hbar v_2 k_+ \\
\hbar v_2 k_-  & \hbar v_3 k_+ & 0             & \hbar v_1 k_- \\
\gamma_1       & \hbar v_2 k_- & \hbar v_1 k_+ & \Delta_0
\end{pmatrix},
\end{equation}
which is in the basis of $A_1$, $B_1$, $A_2$ and $B_2$ sites [Fig.~\ref{Fig:moire}(a)] from one  AB bilayer graphene. $k_{\pm}$ stands for $k_x \pm i k_y$.  Parameter values are taken as 
\begin{equation}
\begin{aligned}
& (v_1, v_2, v_3) = (0.844, -0.045, -0.091)\times 10^6 \ \text{m/s},\\
& \gamma_1 = 361 \ \text{meV}, \Delta_0 = 15 \ \text{meV},
\end{aligned}
\end{equation}
which are extracted from {\it ab initio} results of Ref.~\onlinecite{Jung2014bg}. The moir\'e Hamiltonian in $+K$ valley is given by
\begin{equation}
\mathcal{H}_{+K} = \begin{pmatrix}
h_{\mathfrak{b}}(\kk) & \tilde{T}(\rr) \\
\tilde{T}^{\dagger}(\rr) & h_{ \mathfrak{t}}(\kk)
\end{pmatrix},
\end{equation}
where $h_{\mathfrak{b}}(\kk)$ and $h_{\mathfrak{t}}(\kk)$ are $k.p$ Hamiltonians for bottom and top bilayer graphene, and are equal to $\mathcal{H}_0[\hat{R}(+\theta/2)(\kk-\boldsymbol{\kappa}_+)]$ and $\mathcal{H}_0[\hat{R}(-\theta/2)(\kk-\boldsymbol{\kappa}_-)]$, respectively. Here $\hat{R}(\pm \theta/2)$ are rotation matrices and $\boldsymbol{\kappa}_\pm = [4\pi/(3 a_M)](-\sqrt{3}/2, \mp 1 /2)$. The moir\'e period $a_M$ is approximately $a_0/\theta$, where $a_0$ is the monolayer graphene lattice constant. $\tilde{T}(\rr)$ is the tunneling between bottom and top bilayer graphene, which varies spatially with the moir\'e period as specified by
\begin{equation}
\begin{aligned}
\tilde{T}(\rr)&=\begin{pmatrix}
0 & T(\rr) \\
0 & 0
\end{pmatrix},\\
T(\rr) &= w_0 T_0 +w_1 (e^{-i \bb_+ \cdot \rr} T_{+1}
+e^{-i \bb_- \cdot \rr} T_{-1}), \\
T_j &= \sigma_0 + \cos(2\pi j/3)\sigma_x+\sin(2\pi j/3)\sigma_y
\end{aligned}
\end{equation}
where we only keep tunneling terms between adjacent layers, and $\bb_{\pm}$ are  moir\'e reciprocal lattice vectors
given by $[4\pi/(\sqrt{3} a_M)](\pm1/2, \sqrt{3}/2)$. $w_0$ and $w_1$ are two tunneling parameters, which in general have different numerical values due to layer corrugation in the moir\'e pattern \cite{Koshino2018,choi2019intrinsic}. We take $w_0 = 88$ meV and and $w_1 = 100$ meV.
An out-of-plane electric displacement field generates a layer dependent potential, which can be parametrized using a single parameter $U$ as illustrated in Fig.\ref{Fig:moire}(b). The point group symmetry of TDBG is $D_3$ in the absence of the displacement field ($U=0$), and is broken down to $C_3$ when $U$ is finite.

\begin{figure}[t!]
	\includegraphics[width=1\columnwidth]{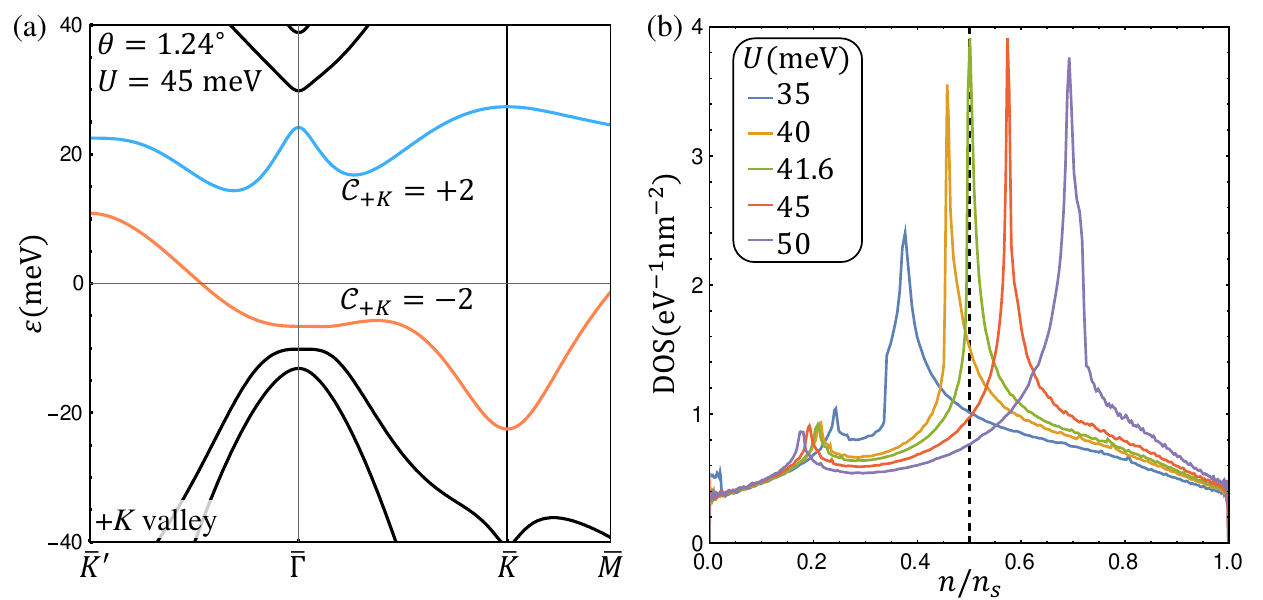}
	\caption{(a) Moir\'e bands in $+K$ valley along high-symmetry lines. The blue (red) band is the first conduction (valence) band above (below) charge neutrality point, and has a Chern number of $+2$ ($-2$) in $+K$ valley. (b) Non-interacting density of states (DOS) per spin and per valley as a function of filling factor. The DOS are tunable by the layer-dependent potential $U$, and peaks at half filling when $U\approx 41.6$ meV. $\theta$ is $1.24^{\circ}$ for numerical studies presented in this paper.}
	\label{Fig:band}
\end{figure}

A representative moir\'e band structure is shown in Fig.~\ref{Fig:band} for $\theta=1.24^{\circ}$ and $U=45$ meV. The first conduction band in Fig.~\ref{Fig:band}(a) is isolated in energy from other bands, narrow in bandwidth($\sim$ 13 meV), and  topologically nontrivial with a Chern number of $+2$ in $+K$ valley. Because of time-reversal symmetry, the corresponding moir\'e band in $-K$ valley has the opposite Chern number.

\begin{figure*}[t!]
	\includegraphics[width=1.6\columnwidth]{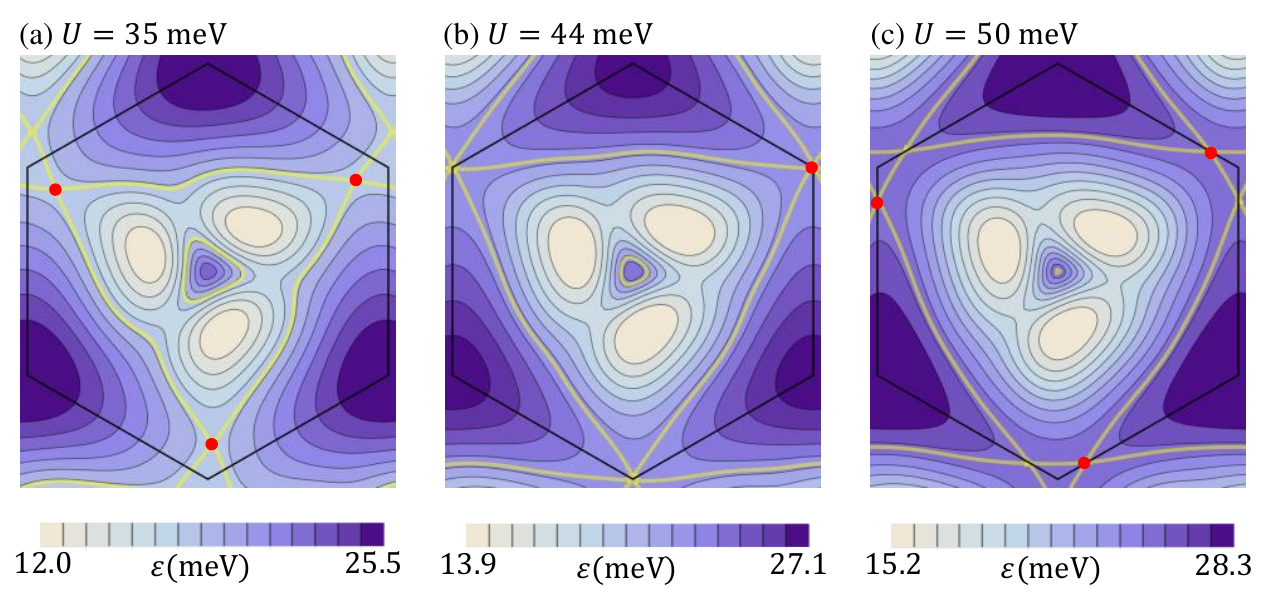}
	\caption{Energy contours for the first moir\'e conduction band in $+K$ valley for different values of $U$. Yellows lines mark  Fermi surface at the van Hove energy. Red points mark inequivalent van Hove saddle points, which merge to one higher-order saddle point in (b). }
	\label{Fig:band2D}
\end{figure*}

We find that the band dispersion can be drastically controlled by the potential $U$, as demonstrated by the energy contour plots of the first moir\'e conduction  band shown in Fig.~\ref{Fig:band2D}. In particular, the van Hove saddle points can be effectively moved in the momentum space by tuning $U$. At a critical $U\approx 44$ meV, three van Hove saddle points merge to the corner of the moir\'e Brillouin zone (MBZ), forming a high-order saddle point \cite{yuan2019magic}.  Correspondingly, the density of states (DOS) for the non-interacting band has a strong dependence on $U$, and the van Hove singularity in the DOS can be tuned from below to above half filling by varying $U$ [Fig.~\ref{Fig:band}(b)]. This strong dependence of the  moir\'e bands on $U$ has implications on interaction physics, as we explain in the following.

%For a given valley, the conduction band minima are located at three inequivalent momenta related by $C_3$, which can explain the degeneracy of 3 for the first Landau fan on the conduction band side \cite{liu2019spin}. Such a moir\'e band with multiple degenerate band extrema can provide a system to study quantum Hall nematic and ferroelectric states in a strong magnetic field \cite{Feldman316,Sodemann2017,Randeria2018}. We note that our parameter values used in the moir\'e Hamiltonian are the same as Ref.~\onlinecite{lee2019theory}, where a systematic study of the moir\'e band structure as a function of $\theta$ and $U$ can be found. In our work, we take the first conduction band in Fig.~\ref{Fig:band}(a) as a concrete example, and study many-body interaction effects within this band.

\section{Ferromagnetism}
\label{section:FM}

\subsection{Ferromagnetic Ground State}

Many-body interactions are effectively enhanced for electrons in the nearly flat moir\'e bands. Here we study flatband ferromagnetism driven by Coulomb repulsion using a momentum-space formalism \cite{Zhang2019, lee2019theory,chen2019tunable}. We only retain the first conduction band for simplicity, and the single-particle Hamiltonian projected onto this band is
\begin{equation}
H_0= \sum_{\kk, \tau, s} \varepsilon_{\kk,\tau} c^{\dagger}_{\kk,\tau, s} c_{\kk,\tau, s},
\end{equation}
where $\kk$ is momentum measured relative to the center of the MBZ, $\tau=\pm$ is the valley index, $s$ represents spin ($\uparrow$, $\downarrow$) and $c^{\dagger}_{\kk,\tau, s}$ is the fermion creation operator. $\varepsilon_{\kk,\tau}$ is the spin independent moir\'e band energy; its valley dependence is determined by time reversal symmetry, and $\varepsilon_{\kk,\tau}=\varepsilon_{-\kk,-\tau}$.

We project Coulomb interaction onto the first conduction band, and the interacting Hamiltonian has the form
\begin{equation}
\begin{aligned}
&H_1 = \frac{1}{2\mathcal{A}}\sum V_{\kk_1 \kk_2 \kk_3 \kk_4}^{(\tau \tau')}
 c^{\dagger}_{\kk_1,\tau, s} c^{\dagger}_{\kk_2,\tau', s'} c_{\kk_3,\tau', s'} c_{\kk_4,\tau, s}, \\
&V_{\kk_1 \kk_2 \kk_3 \kk_4}^{(\tau \tau')} = \sum_{\qq} V(\qq) O_{\kk_1 \kk_4}^{(\tau)} (\qq) O_{\kk_2 \kk_3}^{(\tau')}(-\qq), \\
&O_{\kk_1 \kk_4}^{(\tau)} (\qq) = \sum_{\sigma,\ell} \int d\rr e^{i \qq\cdot \rr} \Phi_{\tau,\kk_1,\sigma,\ell}^{*}(\rr) \Phi_{\tau,\kk_4,\sigma,\ell}(\rr),
\end{aligned}
\end{equation}
where $\mathcal{A}$ is the system area and $\Phi_{\tau,\kk}(\rr)$ is the Bloch wave function for the first conduction band in valley $\tau K$ and at momentum $\kk$. The indices $\sigma$ and $\ell$ respectively label sublattices and layers. By time-reversal symmetry, $\Phi_{\tau,\kk}(\rr)=\Phi^*_{-\tau,-\kk}(\rr)$. In the plane wave matrix element $O_{\kk_1 \kk_4}^{(\tau)} (\qq)$, the momentum $\qq$ can differ from $\kk_1-\kk_4$ by moir\'e reciprocal lattice vectors.  Hamiltonian $H_1$ represents density-density interaction, and preserves spin SU(2) and valley U(1) symmetry. In fact, $H_1$ has an enlarged SU(2)$\times$SU(2) symmetry, which stands for an independent spin rotational symmetry within each valley. Short-range interactions (e.g., atomic scale on-site Hubbard repulsion), which we do not study explicitly, breaks the SU(2)$\times$SU(2) symmetry down to spin SU(2) symmetry.

The Coulomb interaction $V(\qq)$ can be screened by dielectric environment and nearby metallic gates.  We  assume that TDBG  is encapsulated by an insulator (typically boron nitride), and is in the middle to two metallic gates, which generate an infinite series of equally spaced image charges with alternating signs. Under this image charge approximation, the screened Coulomb potential in momentum space is 
$V(\qq)= 2\pi e^2\tanh(q d)/(\epsilon q)$,
where $\epsilon$ is the dielectric constant of the encapsulating insulator, and $d$ is the vertical distance between the top (bottom) metallic gate and TDBG. We take $d$ to be 50 nm for all calculations presented in the following.   The Coulomb interaction energy scale is set by $E_C=e^2/(\epsilon a_M)$. At $\theta=1.24^{\circ}$, $a_M\approx 11.4$nm, and $E_C\approx$12 meV for $\epsilon=10$. Since the typical Coulomb interaction energy scale is comparable to the bandwidth ($\sim$ 10 meV), there is a strong tendency towards symmetry-breaking phases driven by interactions. The system is characterized by almost-flat narrow noninteracting bands with large Coulomb energy, a classic situation for the manifestation of strong correlation physics.

We use Hartree-Fock (HF) approximation and assume that both moir\'e periodicity and valley U(1) symmetry are preserved, but allow spin polarization, motivated by the experimental evidence of ferromagnetism \cite{shen2019observation,liu2019spin,cao2019electric} in TDBG. This leads to the following mean-field HF Hamiltonian
\begin{equation}
\begin{aligned}
H_{\text{MF}} =& \sum_{\kk, \tau, s} E_{\kk,\tau, s} c^{\dagger}_{\kk,\tau, s} c_{\kk,\tau, s},\\
E_{\kk,\tau, s} =& \varepsilon_{\kk,\tau}+\Sigma_{\kk,\tau,s},\\
\Sigma_{\kk,\tau,s} =&\frac{1}{A} \sum_{\kk', \tau', s'} V_{\kk \kk' \kk' \kk}^{(\tau \tau')} n_F(E_{\kk',\tau', s'})\\
&-\frac{1}{A} \sum_{\kk'} V_{\kk \kk' \kk \kk'}^{(\tau \tau)} n_F(E_{\kk',\tau, s})
\end{aligned}
\label{HFM}
\end{equation}
where the quasiparticle energy $E_{\kk,\tau, s}$ includes moir\'e band energy $\varepsilon_{\kk,\tau}$ and Hartree-Fock self energy $\Sigma_{\kk,\tau,s}$, and $n_F$ is the Fermi-Dirac occupation number.  By projecting the interaction onto the first conduction band, we  neglected self-energy induced by interaction with all other occupied bands. This is expected to be a reasonable approximation for TDBG because of the energetic separation of neighboring bands from the first conduction band.  The neighboring band effects could, in principle, be included in the theory, if necessary, either by developing a brute-force multiband mean field theory or perturbatively, with the higher bands contributing paramterically smaller terms because of the large energy denominators associated with band separations.

\begin{figure}[t!]
	\includegraphics[width=0.7\columnwidth]{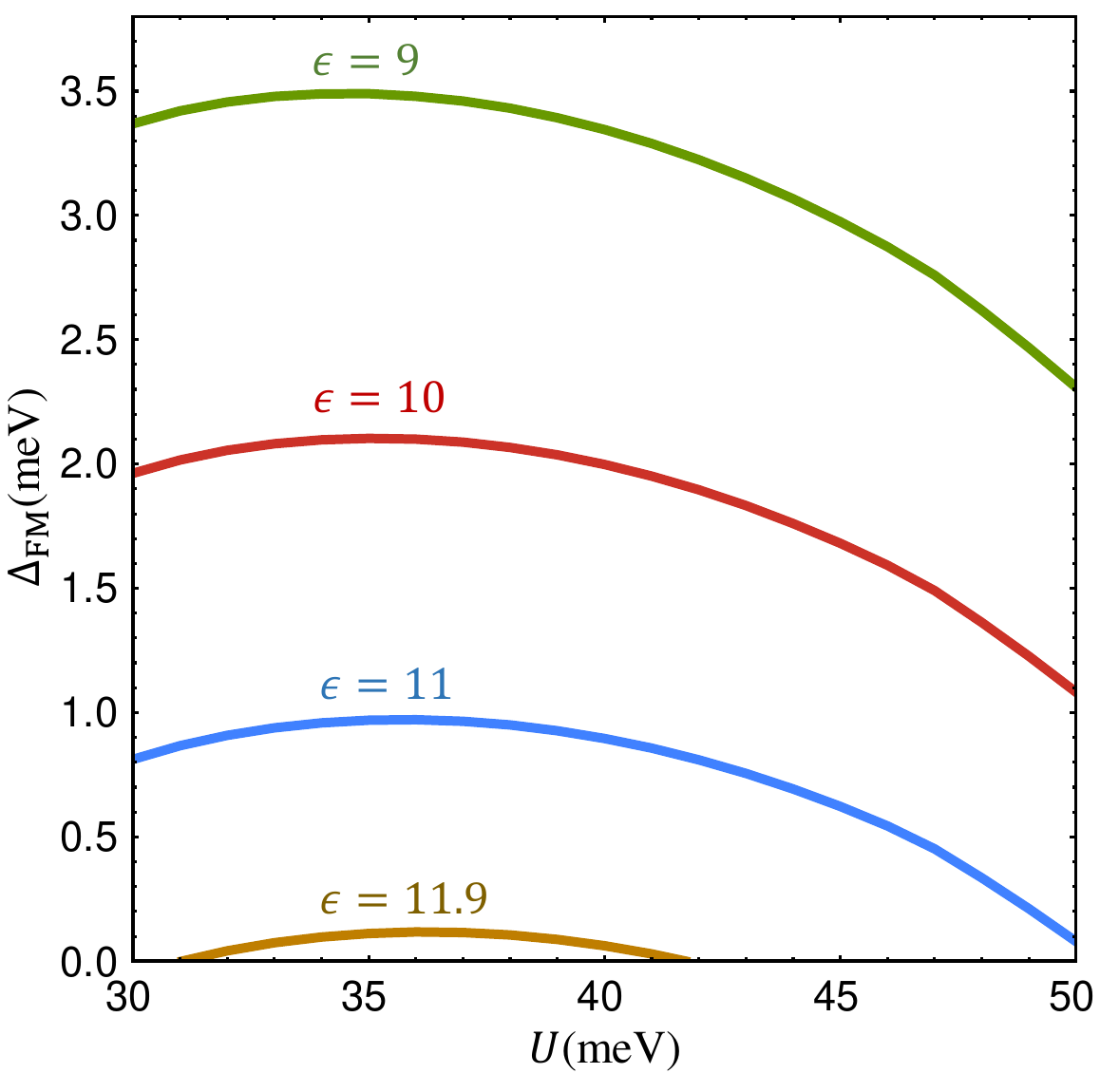}
	\caption{The ferromagnetic insulating gap $\Delta_{\text{FM}}$ at half filling as a function of the layer potential difference $U$ for different values of dielectric constant $\epsilon$. $\Delta_{\text{FM}}$  has a dome shape, which explains the experimental observation \cite{shen2019observation,liu2019spin,cao2019electric,Burg2019,he2020tunable} that the FMCI appear only over a finite range of displacement field. }
	\label{Fig:gap}
\end{figure}

We denote the electron filling factor as $\nu=n/n_s$, where $n$ is the electron density, and $n_s$ the density for 4 electrons per moir\'e cell. At half filling $\nu=1/2$ that corresponds to 2 electrons per moir\'e cell, the mean-field theory in Eq.~(\ref{HFM}) leads to two distinct symmetry-broken states, namely, valley-polarized state and valley-unpolarized state, which are degenerate at this particular filling.  Short-range interactions can break this symmetry, as explained in Appendix~\ref{appA}. Because the moir\'e  conduction bands carry a valley-contrast Chern number, the valley polarized state supports quantum anomalous Hall effect (QAHE). To our knowledge, QAHE has not yet been observed in TDBG. Therefore, we leave the valley polarized state at $\nu=1/2$ in TDBG to future study, and focus on valley unpolarized states in the following. 

Because the Hamiltonian $H=H_0+H_1$ has the enlarged SU(2)$\times$SU(2) symmetry, a valley unpolarized state can have independent spin polarization in the two valleys. For example, the two valleys can have either parallel or antiparallel spin polarization. However, atomic-scale on-site Hubbard interaction explicitly breaks SU(2)$\times$SU(2) down to SU(2) symmetry, and selects the ferromagnetic state in which spins of the two valleys are polarized along the same direction (see Appendix~\ref{appA}). In the following, we only consider valley unpolarized state with identical spin polarization in the two valleys. Therefore, we make the ansatz that $\Sigma_{\kk,\tau, s}=\Sigma_{-\kk,-\tau, s}$. This mean-field ansatz preserves spinless time-reversal symmetry.

With the above ansatz, we solve the mean-field theory in Eq.~(\ref{HFM}) self consistently. One characteristic quantity is the zero-temperature ($T=0$) ferromagnetic gap $\Delta_{\text{FM}}$ that separates the occupied spin majority states from the unoccupied spin minority states at  $\nu=1/2$. We plot $\Delta_{\text{FM}}$ as a function of the layer dependent potential $U$ for different dielectric constant $\epsilon$ in Fig.~\ref{Fig:gap}, which shows that $\Delta_{\text{FM}}$ has a dome shape and is positive only over a finite range of $U$ for large $\epsilon$. The dome shape in $\Delta_{\text{FM}}$  correlates with the non-interacting DOS [Fig.~\ref{Fig:band}(b)], which peaks at $\nu=1/2$ at $U\approx 41 $meV. A larger non-interacting DOS at $\nu=1/2$ implies a stronger instability towards symmetry breaking, and therefore, a larger interaction driven energy gap. However, we note that this argument is only qualitative, as $\Delta_{\text{FM}}$ does not exactly follow the non-interacting DOS. An important conclusion we can draw from Fig.~\ref{Fig:gap} is that the FM insulating gap is tunable by an external displacement field, which agrees with the experimental observation that the FMCI at $\nu=1/2$ only develops over a finite range of displacement field \cite{shen2019observation,liu2019spin,cao2019electric,Burg2019,he2020tunable}.

\begin{figure}[t!]
	\includegraphics[width=1\columnwidth]{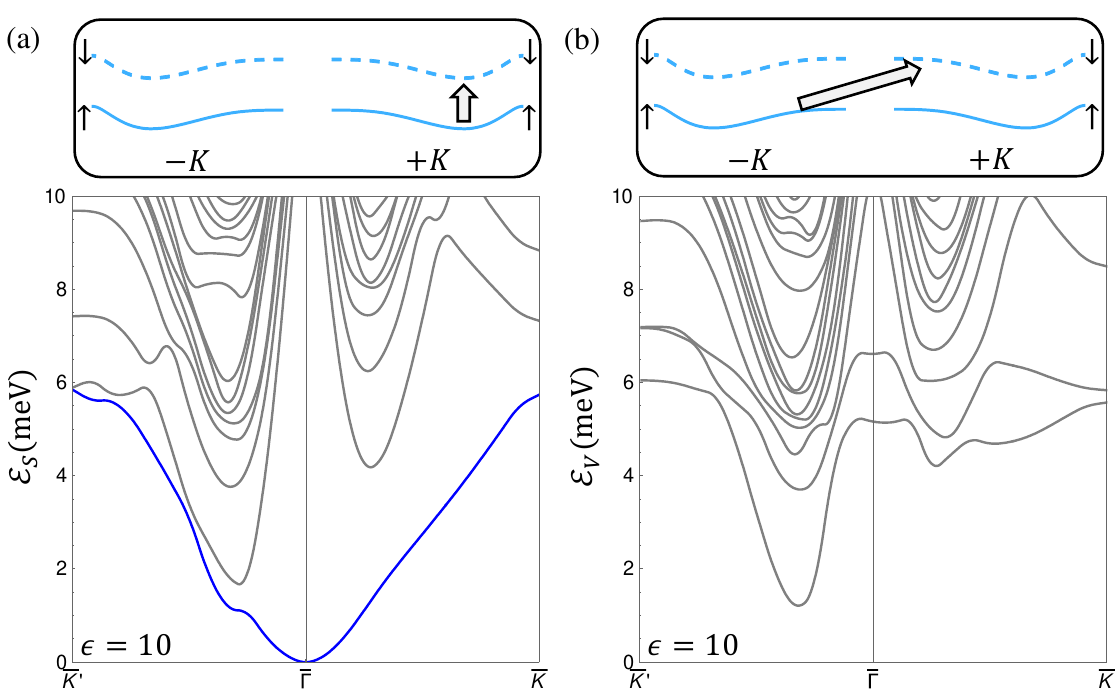}
	\caption{(a) Excitation spectrum for  spin magnons (illustrated in the upper panel). The blue line marks the gapless spin wave mode. (b)  Excitation spectrum for  valley magnons (illustrated in the upper panel).  The valley magnon spectrum is fully gapped. $U$ is 45 meV for the calculations. }
	\label{Fig:wave}
\end{figure}

\subsection{Magnon Spectrum}
A positive FM insulating gap $\Delta_{\text{FM}}$ indicates that the FM state is a good ansatz at the mean-field level. 
To examine whether the FM state is stable beyond mean-field theory, we calculate the energy spectrum for one-magnon collective excitations. There are two types of magnons for the FM insulator at half filling, namely, spin magnons and valley magnons. The spin magnons involve collective particle-hole transitions from the occupied spin majority band to the unoccupied spin minority band within the same valley, as illustrated in Fig.~\ref{Fig:wave}(a);  the valley magnons are collective particle-hole transitions that flip the valley index, as shown  in Fig.~\ref{Fig:wave}(b). We calculate the spin and valley magnon spectrum separately by solving their corresponding Bethe-Salpeter equations, following the theory developed in Ref.~\onlinecite{Wu2020Collective}. Details of the theory can also be found in Appendix~\ref{appB}.

Representative spectra for spin and valley magnons are shown in Figs.~\ref{Fig:wave}(a) and \ref{Fig:wave}(b), respectively. The spin magnon spectrum has gapless spin wave modes, consistent with the Goldstone's theorem, as the continuous SU(2)$\times$SU(2) symmetry is spontaneously broken in the ferromagnetic state. In fact, the SU(2) symmetry associated with each valley is broken, so there are two spin wave modes, one for each valley. The overall spin excitation
spectrum is nonnegative in Fig.~\ref{Fig:wave}(a), showing the stability of the FM insulator against spin-magnon excitations.

By contrast, the valley magnon spectrum shown in Fig.~\ref{Fig:wave}(b) is gapped. This is consistent with the fact that there is only U(1) symmetry in the valley space, and the FM insulator does not break this valley U(1) symmetry. The positive valley magnon spectrum indicates the stability of the FM insulator against valley-magnon excitations, and also implies that the FM insulator is energetically more favorable than inter-valley density wave state.

Based on the spectrum shown in Fig.~\ref{Fig:wave}, we conclude that the half-filled  FMCI in TDBG can be stable against one-magnon collective excitations.

\subsection{Mean-Field Transition Temperature}

We now turn to finite temperature physics and calculate the mean-field transition temperature $T_{\text{FM}}$ for the FM phase. To determine $T_{\text{FM}}$, we define
$\Sigma_{\kk,\tau}^{(\pm)} = (\Sigma_{\kk,\tau,\uparrow} \pm \Sigma_{\kk,\tau,\downarrow})/2$. At $T_{\text{FM}}$, $\Sigma_{\kk,\tau}^{(-)}$ is infinitesimally small, and the self-consistent equation (\ref{HFM}) can be linearized as follows
\begin{equation}
\begin{aligned}
\Sigma_{\kk,\tau}^{(-)} & = \sum_{\kk'} M_{\kk \kk'}^{(\tau)} \Sigma_{\kk',\tau}^{(-)},\\
M_{\kk \kk'}^{(\tau)} & = -\frac{1}{\mathcal{A}} V_{\kk \kk' \kk \kk'}^{(\tau \tau)} \frac{\partial n_F(E)}{\partial E}|_{E=\varepsilon_{\kk'}+\Sigma_{\kk',\tau}^{(+)}},
\end{aligned}
\end{equation}
from which $T_{\text{FM}}$ can be obtained by requiring the largest eigenvalue of the matrix  $M^{(\tau)}$ to be 1.

\begin{figure}[t!]
	\includegraphics[width=0.8\columnwidth]{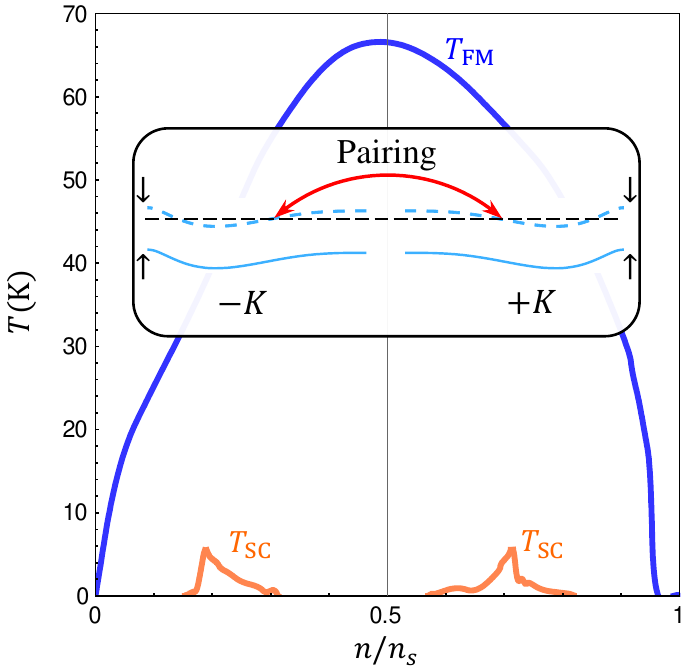}
	\caption{Mean-field transition temperatures for ferromagnetism (blue lines) and superconductivity (red lines) as a function of filling factor $n/n_s$. $U$ is 45 meV and $\epsilon$ is 11 for the calculation. The inset is a schematic illustration for the inter-valley equal-spin pairing in the ferromagnetic state. } 
	\label{Fig:Tc}
\end{figure}

A representative plot of $T_{\text{FM}}$ as a function of filling factor is shown in Fig.~\ref{Fig:Tc}, where  ferromagnetism develops over a large range of filling factors with $T_{\text{FM}}$ up to few tens of kelvin. We note that our mean-field theory overestimates the tendency towards ordering, as fluctuations like spin waves are neglected in the estimation of $T_{\text{FM}}$.  The ferromagnetic state at half filling can be an insulator at zero temperature, when the spin  majority bands are fully filled and separated from the empty  spin minority bands by an energy gap $\Delta_{\text{FM}}$. Away from half filling,  the ferromagnetic state is generically metallic with spin dependent Fermi surfaces.

\section{Superconductivity}
\label{section:SC}

The metallic state away from half filling can be susceptible to superconducting instability due to enhanced electron-phonon interaction in moir\'e flatband systems. Here we study superconductivity mediated by electron-acoustic phonon interactions. The in-plane acoustic longitudinal phonon modes mediate effective electron attraction as follows
\begin{equation}
H_{\text{att}}= -g_0 \sum_{\sigma,\sigma',\ell,s}\int d\rr \hat{\psi}_{+ \sigma \ell s}^\dagger  \hat{\psi}_{- \sigma' \ell s'}^\dagger
\hat{\psi}_{- \sigma' \ell s'} \hat{\psi}_{+ \sigma \ell s},
\label{Hatt}
\end{equation}
where $\hat{\psi}_{\tau \sigma \ell s}(\rr)$ is the electron field operator at the coarse-grained position $\rr$ associated with valley $\tau K$, sublattice $\sigma=A, B$, layer $\ell=1,2,3,4$ and spin $s=\uparrow,\downarrow$. In Eq.~(\ref{Hatt}), we only retain attractive interactions that pair electrons from opposite valleys. The coupling constant $g_0$ is given by $D^2/(\rho_m v_{s}^2)$, where $D$ is the deformation potential, $\rho_m$ is the mass density of monolayer graphene, and $v_s$ is the velocity of acoustic longitudinal phonon. Using $D= 30$ eV, $\rho_m=7.6\times 10^{-8}$ g/cm${^2}$, $v_s=2\times 10^6$ cm/s, we estimate $g_0$ to be 474  meV nm$^2$. Here we neglect retardation effects in the phonon mediated electron attraction for simplicity. 

As we showed previously in Ref.~\onlinecite{wu2019phonon}, the attraction in Eq.(\ref{Hatt}) can be decomposed into four different pairing channels that are distinguished by their orbital and spin characters: 
(1)  intrasublattice spin-singlet $s$-wave pairing, i.e., $(is_y)_{s s'} \hat{\psi}_{+ \sigma \ell s}^\dagger  \hat{\psi}_{-  \sigma \ell s'}^\dagger $; 
(2)  intersublattice spin-triplet $p$-wave pairing, e.g., $\mathcal{F}_{s s'} \hat{\psi}_{+ A \ell s}^\dagger  \hat{\psi}_{- B \ell s'}^\dagger $, where $\mathcal{F}$ can be any one of the three symmetric tensors $(s_0 \pm s_z)/2$ and  $s_x$; 
(3)  intersublattice spin-singlet $d$-wave  pairing, e.g., $(is_y)_{s s'} \hat{\psi}_{+ A \ell s}^\dagger  \hat{\psi}_{- B \ell s'}^\dagger $; 
and (4) intrasublattice spin-triplet  $f$-wave pairing, i.e., $\mathcal{F}_{s s'} \hat{\psi}_{+ \sigma \ell s}^\dagger  \hat{\psi}_{- \sigma \ell s'}^\dagger $.
The $s$-wave and $f$-wave pairings are  only distinguished by their spin characters, and the same is true for $p$ and $d$ pairings. The angular momenta of intersublattice Cooper pairs arise from the valley-contrast sublattice chirality under $C_3$ rotation \cite{Wu2018phonon, wu2019phonon}.  In $AB$ bilayer graphene, one of the sublattices in each layer [$A_1$ and $B_2$ sites in Fig.~\ref{Fig:moire}(b)] is pushed to higher energy by interlayer tunneling. Therefore, intersublattice pairing is energetically less favorable compared to intrasublattice pairing in TDBG. In the following, we only consider interactions that pair electrons on the same sublattice, and project such interactions onto the first moir\'e conduction band. The projected pairing Hamiltonian is
\begin{equation}
\begin{aligned}
H_{\text{p}}&=-\frac{1}{\mathcal{A}}\sum g_{\kk \kk'} c_{\kk,+,s}^{\dagger} c_{-\kk,-,s'}^{\dagger} c_{-\kk',-,s'} c_{\kk',+,s},\\
g_{\kk \kk'}&= g_0 \mathcal{A} \sum_{\sigma,\ell} \int d\rr |\Phi_{+,\kk,\sigma,\ell}(\rr)|^2 |\Phi_{+,\kk',\sigma,\ell}(\rr)|^2,
\end{aligned}
\end{equation}
where we only keep interactions that pair electrons with opposite momenta, i.e., momentum $\kk$ in $+K$ valley and momentum $-\kk$ in $-K$ valley. The pairing Hamiltonian $H_{\text{p}}$ also has the SU(2)$\times$SU(2) symmetry, and supports both spin singlet $s$-wave and spin triplet $f$-wave pairings. Because of the ferromagnetism induced by Coulomb repulsion, equal spin pairing is more favored compared to spin singlet pairing. Therefore, we consider intervalley pairing between electrons with the same spin, which leads to the following Bardeen-Cooper-Schrieffer (BCS) mean-field Hamiltonian
\begin{equation}
\begin{aligned}
H_{\text{BCS}} &= -\sum_{\kk,s} ( \Delta_{\kk,s}  c_{\kk,+,s}^{\dagger} c_{-\kk,-,s}^{\dagger} +\text{H.c.}),\\
\Delta_{\kk,s} &= \frac{1}{\mathcal{A}}\sum_{\kk'} g_{\kk \kk'} \langle c_{-\kk',-,s} c_{\kk',+,s} \rangle.
\end{aligned}
\end{equation}

By combining the BCS Hamiltonian $H_{\text{BCS}} $ and the effective single-particle Hamiltonian $H_{\text{FM}}$ [Eq.~(\ref{HFM})] that is renormalized by the Coulomb interaction, we obtain the superconducting linearized gap equation 
\begin{equation}
\begin{aligned}
\Delta_{\kk,s} &= \sum_{\kk'} \chi_{\kk \kk'}^{(s)} \Delta_{\kk',s},\\
\chi_{\kk \kk'}^{(s)} &= \frac{g_{\kk \kk'} }{\mathcal{A}} \frac{1-2n_F(E_{\kk',+,s})}{2 (E_{\kk',+,s}-\mu)},
\end{aligned}
\label{chiSC}
\end{equation}
where $\mu$ is the chemical potential, and $E_{\kk,\tau,s}=\varepsilon_{\kk,\tau}+\Sigma_{\kk,\tau,s}$ is the effective band energy including the self energy. We have used the spinless time-reversal symmetry, which implies $E_{\kk,\tau,s}=E_{-\kk,-\tau,s}$,  to simplify the superconducting susceptibility $\chi^{(s)}$. Because of this symmetry, ferromagnetism does not lead to depairing effect for superconductivity with intervalley equal-spin pairing. In Eq.~(\ref{chiSC}), spin up  and down channels have independent gap equations. The superconducting transition temperature $T_{\text{SC}}$ is reached when the largest eigenvalue of $\chi^{(s)}$ is 1. Fig.~\ref{Fig:Tc} plots $T_{\text{SC}}$ as a function of filling factor, and shows two superconducting domes respectively on the two sides of the half-filled ferromagnetic state. In Fig.~\ref{Fig:Tc},  we take the value of $g_0$ (the attractive interaction strength) to be three times of   474  meV nm$^2$ (the value obtained from the above electron-acoustic phonon coupling parameters) in order to get a value of $T_{\text{SC}}$ on the order of 1 K. We note that $T_{\text{SC}}$ is exponentially sensitive to $g_0$ as well as the moir\'e band flatness. A quantitative study of $T_{\text{SC}}$ is beyond the scope of this paper. In any case, the experimental parameters are not known with sufficient accuracy for a quantitative estimate of $T_{\text{SC}}$ at this stage of development of the field. The main purpose of this section is to point out the possibility of phonon-mediated spin triplet pairing in a ferromagnetic system.

We discuss the effect of an in-plane magnetic field $\boldsymbol{B}_{\parallel}$ on  $T_{\text{SC}}$ in the $f$-wave channel. If the parent state for superconductivity is spin unpolarized, then $T_{\text{SC}}$ can be slightly enhanced by $\boldsymbol{B}_{\parallel}$ in the low-field regime, because Zeeman energy leads to  an effective spin dependent chemical potential shift \cite{lee2019theory, wu2019identification,scheurer2019pairing}. On the other hand, if the parent state already has maximum spin polarization allowed by a given filling factor, then an externally applied $\boldsymbol{B}_{\parallel}$ field can no longer change the amount of spin polarization, and  $T_{\text{SC}}$ is reduced by $\boldsymbol{B}_{\parallel}$ due to orbital effect \cite{lee2019theory, wu2019identification}.  In Ref.\onlinecite{liu2019spin},  $T_{\text{SC}}$ is found  to be slightly enhanced by weak $\boldsymbol{B}_{\parallel}$ field, indicating that the superconducting state has spin triplet pairing but with no spin polarization. Our mean-field phase diagram in Fig.~\ref{Fig:Tc} likely overestimates the filling range for ferromagnetism. We emphasize that there is always a superconducting instability in a partially filled band regardless of the presence or absence of ferromagnetism in our theory, where the superconductivity is mediated by electron-phonon interactions and ferromagnetism is driven by Coulomb repulsion. An additional signature of electron-acoustic phonon interaction in TDBG is that phonon scattering can lead to large linear-in-$T$ resistivity in transport above some crossover temperatures \cite{li2019phonon}. 

Finally, We note that experimental signatures of SC in TDBG are not yet conclusive, as discussed in detail in Ref.~\onlinecite{he2020tunable}.

\section{Conclusion}
\label{section:end}

In conclusion, we have presented a theory of ferromagnetism induced by Coulomb repulsion and superconductivity mediated by electron-acoustic phonon interactions in moir\'e bands of TDBG. In our theoretical phase diagram, there can be a ferromagnetic correlated insulator at half filling, and  superconducting domes on both the electron and hole sides of the half-filled insulator. %The resemblance between our theoretical phase diagram and the experimental phase diagram of TDBG \cite{shen2019observation,liu2019spin, cao2019electric} in some key aspects is encouraging. 
Ferromagnetism  and superconductivity are two prototypical orders that can occur in moir\'e flat bands, while there are many other possible competing and/or intertwined orders, such as nematicity that breaks rotational symmetry and density wave state that breaks moir\'e translation symmetry \cite{hsu2020topological}. In TDBG, there is experimental evidence that states with both spin and valley polarization are possibly stabilized at 1/4 and 3/4 fillings by a finite in-plane magnetic field \cite{liu2019spin, cao2019electric, he2020tunable}. CIs at these factors could also be spin and/or valley polarized states. Because of the valley contrast Chern numbers in the non-interacting moir\'e bands, valley polarized CIs can also display quantum anomalous Hall effects.  Our work should be viewed as a step towards a full quantitative theory of the potentially very rich TDBG phase diagram. A note-worthy qualitative feature of the current work is the possibility, already apparent at the mean field level, that SC and FMCI phases, although they arise from different interactions (electron-phonon for SC and electron-electron for FMCI), could compete with each other in TDBG moir\'e flatband with the FM phase centered around half-filling and the SC domes manifesting on both electron- and hole-doped sides of half-filling.  The fact that this could be the experimental TDBG situation may indicate that our theory captures some essential qualitative aspect of moir\'e interaction physics although our use of mean field theory (and many other approximations, e.g., neglect of higher bands) exaggerates the quantitative stability of the symmetry-broken phases compared with experiments.

TDBG and other related moir\'e systems represent a highly tunable platform, where moir\'e band structure can be effectively controlled by the out-of-plane displacement field, as revealed by our theoretical study. This feature allows {\it in situ}  control of band structure, and provides unprecedented opportunities to study many-body physics.

\section{acknowledgment}
F. W. thanks Y.-T. Hsu,  X. Li, and R.-X. Zhang for discussions. This work is supported by  Laboratory for Physical Sciences.

\appendix
\section{Short-Range Interactions}
\label{appA}

We show that short-range interactions, in particular, the atomic scale on-site Hubbard repulsion, explicitly break the SU(2)$\times$SU(2) symmetry down to spin SU(2) symmetry, and favors ferromagnetic states in which spins in the two valleys are polarized along the same direction.  

The on-site Hubbard repulsion on  monolayer graphene honeycomb lattice is given by
\begin{equation}
\begin{aligned}
H_2 &= U_0 \sum_{\RR, \sigma} b^{\dagger}_{\RR\sigma \uparrow}b^{\dagger}_{\RR\sigma \downarrow}b_{\RR\sigma\downarrow}b_{\RR\sigma\uparrow}\\
&=\frac{U_0}{N}\sum_{\kk_i}^{'} \sum_{\sigma}  b^{\dagger}_{\kk_1\sigma \uparrow}b^{\dagger}_{\kk_2\sigma \downarrow}b_{\kk_3\sigma\downarrow}b_{\kk_4\sigma\uparrow},
\end{aligned}
\label{H0}
\end{equation}
where $U_0$ is the on-site Hubbard interaction, $\RR$ is the lattice vector, $\sigma$ is the sublattice index ($A$, $B$), $N$ is the number of unit cells in the monolayer, and the prime on the summation of the second line is the momentum conservation constraint, i.e., $\kk_1+\kk_2$ is equivalent to $\kk_3+\kk_4$ modulo reciprocal lattice vectors. To obtain a continuum model, we only keep  states near $\pm K$ points:
\begin{equation}
H_2\approx \frac{U_0}{N}
\sum_{\kk_i}^{'} \sum_{\tau_i}^{'}\sum_{\sigma}
b^{\dagger}_{\kk_1 \tau_1 \sigma \uparrow}b^{\dagger}_{\kk_2 \tau_2 \sigma \downarrow}b_{\kk_3 \tau_3 \sigma\downarrow}b_{\kk_4 \tau_4 \sigma\uparrow},
\end{equation}
where $\tau = \pm 1$ is the valley index, and the prime on the summation of $\tau_i$ implies the valley conservation $\tau_1+\tau_2=\tau_3+\tau_4$ due to momentum conservation. In the operator $b^{\dagger}_{\kk \tau \sigma s}$, the momentum $\kk$ is measured relative to $\tau K$. We make a Fourier transformation to introduce the coarse-grained real-space position $\rr$:
\begin{equation}
b^{\dagger}_{\kk \tau \sigma s} = \frac{1}{ \sqrt{\mathcal{A}} }\int d\rr e^{i \kk\cdot\rr} \hat{\psi}^{\dagger}_{\tau \sigma s}(\rr),
\end{equation}
where $\mathcal{A}$ is the system area.
The on-site repulsion can then be transformed to a continuum Hamiltonian with local interaction:
\begin{equation}
H_2 \approx u_0 \sum_{\tau_i}^{'} \sum_{\sigma} \int d\rr 
\hat{\psi}^{\dagger}_{\tau_1 \sigma \uparrow}
\hat{\psi}^{\dagger}_{\tau_2 \sigma \downarrow}
\hat{\psi}_{\tau_3 \sigma \downarrow}
\hat{\psi}_{\tau_4 \sigma \uparrow},
\label{H_delta}
\end{equation}
where $u_0= U_0 \mathcal{A}_0$, and $\mathcal{A}_0 = \sqrt{3} a_0^2 /2$ is the area per unit cell in the monolayer.  The local repulsion in Eq.~(\ref{H_delta}) can swap the valley indices of a pair of electrons, and therefore, break the SU(2)$\times$SU(2) symmetry down spin SU(2) symmetry.

We project $H_2$ in Eq.~(\ref{H_delta}) to the first moir\'e conduction band,  perform Hartree-Fock decomposition using the ansatz given in the main text, and obtain the following mean-field Hamiltonian:
\begin{equation}
\begin{aligned}
\tilde{H}_{2}=\sum_{\tau, \kk} \big\{
&[\frac{1}{\mathcal{A}}\sum_{\tau' \kk'} u_{\kk \kk'}^{(\tau \tau')} \langle c_{\kk', \tau', \downarrow}^{\dagger}  c_{\kk', \tau', \downarrow} \rangle ] c_{\kk, \tau, \uparrow}^{\dagger}  c_{\kk, \tau, \uparrow} \\
+&[\frac{1}{\mathcal{A}}\sum_{\tau' \kk'} u_{\kk \kk'}^{(\tau \tau')} \langle c_{\kk', \tau', \uparrow}^{\dagger}  c_{\kk', \tau', \uparrow} \rangle ] c_{\kk, \tau, \downarrow}^{\dagger}  c_{\kk, \tau, \downarrow}
\big\}, 
\end{aligned}
\label{HUMF}
\end{equation}
\begin{equation}
u_{\kk \kk'}^{(\tau \tau')} = u_0 \mathcal{A} \sum_{\sigma \ell} \int d\rr |\Phi_{\tau, \kk, \sigma,\ell}(\rr)|^2 |\Phi_{\tau', \kk', \sigma,\ell}(\rr)|^2.
\end{equation}

It is clear from Eq.~(\ref{HUMF}) that the local repulsion favors a valley unpolarized but spin polarized ferromagnetic state with spins associated with the two valley polarized to the same direction.

The energy scale for the short-range repulsion is $E_U=U_0 a_0^2/a_M^2$, which is about 2 meV using $U_0 = 5$ eV, and is an order-of-magnitude weaker compared to the long-range Coulomb interaction.  The atomic-scale on-site Hubbard interaction acts a weak anisotropy that breaks SU(2)$\times$SU(2)  down to SU(2) symmetry, and selects a particular set of states out of an SU(2)$\times$SU(2) multiplet. For examples, the short-range Hubbard interaction aligns spins associated with the two valleys in the ferromagnetic phase, and suppresses $s$-wave but not $f$-wave pairing in the superconducting phase.

\section{Theory for magnon excitations}
\label{appB}

In this Appendix, we present the theory for magnon excitations, which has been discussed in moir\'e systems in Refs.~\onlinecite{Wu2020Collective} and \onlinecite{alavirad2019ferromagnetism}. The spin magnon states can be parametrized as follows
\begin{equation}
|\QQ\rangle_S = \sum_{\kk} z_{\kk,\QQ} c_{\kk+\QQ,+,\downarrow}^{\dagger} c_{\kk,+,\uparrow} |\text{FM}\rangle
\label{spinwave_Q}
\end{equation}
where $|\text{FM}\rangle$ is the half-filled FM insulating state in which spin $\downarrow$ bands in both valleys are empty, $z_{\kk,\QQ}$ are variational parameters, and $\QQ$  is the momentum of the magnon. In the magnon state $|\QQ\rangle_S$, we make a single spin flip from the occupied  spin $\uparrow$  band to unoccupied spin $\downarrow$  band within the same $+K$ valley. Variation of the magnon energy with respect to $z_{\kk,\QQ}$ leads to the following Bethe-Salpeter equation
\begin{equation}
\begin{aligned}
&\mathcal{E}_S(\QQ)z_{\kk,\QQ}=\sum_{\kk'} \mathcal{H}_{\kk \kk'}^{(\QQ)}z_{\kk',\QQ},\\
&\mathcal{H}_{\kk \kk'}^{(\QQ)}= (E_{\kk+\QQ,+,\downarrow}-E_{\kk,+,\uparrow})\delta_{\kk,\kk'}-\frac{1}{A}V^{(++)}_{\kk' (\kk+\QQ) (\kk'+\QQ) \kk},
\end{aligned}
\label{SpinWave}
\end{equation}
where the first part in $\mathcal{H}_{\kk \kk'}^{(\QQ)}$ is the quasiparticle energy cost of the particle-hole transition, and the second part represents the electron-hole attraction. The eigenvalue $\mathcal{E}_S(\QQ)$ represents the energy of spin magnons. We note that there is another spin wave mode in $-K$ valley, which can be formulated in a similar way as Eq.~(\ref{spinwave_Q}).

In addition to spin magnon states, there are also valley magnon states with a valley flip 
\begin{equation}
|\QQ\rangle_V = \sum_{\kk} z_{\kk,\QQ} c_{\kk+\QQ,+,\downarrow}^{\dagger} c_{\kk,-,\uparrow} |\text{FM}\rangle.
\end{equation}
The corresponding Bethe-Salpeter equation  is
\begin{equation}
\begin{aligned}
&\mathcal{E}_V(\QQ)z_{\kk,\QQ}=\sum_{\kk'} \mathcal{W}_{\kk \kk'}^{(\QQ)}z_{\kk',\QQ},\\
&\mathcal{W}_{\kk \kk'}^{(\QQ)}= (E_{\kk+\QQ,+,\downarrow}-E_{\kk,-,\uparrow})\delta_{\kk,\kk'}-\frac{1}{A}V^{(-+)}_{\kk' (\kk+\QQ) (\kk'+\QQ) \kk},
\end{aligned}
\end{equation}
which gives rise to the valley magnon spectrum in Fig.~\ref{Fig:wave}(b).

\bibliographystyle{apsrev4-1}
\bibliography{refs}

\end{document}